\documentclass[sigconf,10pt,nonacm]{acmart}
\microtypesetup{expansion=false}
\setcopyright{none}

\usepackage[utf8]{inputenc}
\usepackage[T1]{fontenc}
\usepackage{hyperref}
\usepackage{url}
\usepackage{booktabs}
\usepackage{amsmath}
\usepackage{amsfonts}
\usepackage{nicefrac}
\usepackage{algorithm}
\usepackage{algpseudocode}
\usepackage{tikz}
\usetikzlibrary{arrows.meta,positioning}
\usepackage{fancyvrb}
\usepackage{xcolor}
\usepackage{listings}
\newcommand\getsdollar{\mathrel{{\leftarrow}\vcenter{\hbox{\tiny\rmfamily\upshape\$}}}}

\title{Whistledown: Combining User-Level Privacy with Conversational Coherence in LLMs}

\author{Chelsea McMurray}
\affiliation{%
  \institution{Dorcha}
}
\email{chelsea@getdorcha.com}

\author{Hayder Tirmazi}
\affiliation{%
  \institution{Dorcha}
}
\email{hayder@getdorcha.com}

\renewcommand{\shortauthors}{C. McMurray and H. Tirmazi}

\begin{document}

\begin{abstract}
Users increasingly rely on large language models (LLMs) for personal, 
emotionally charged, and socially sensitive conversations. However, 
prompts sent to cloud-hosted models can contain personally identifiable 
information (PII) that users do not want logged, retained, or leaked. 
We observe this to be especially acute when users discuss friends, coworkers, or adversaries, i.e., when they \textit{spill the tea}. Enterprises face the same challenge when they want 
to use LLMs for internal communication and decision-making. 

In this whitepaper, we present Whistledown, a best-effort privacy 
layer that modifies prompts \textit{before} they are sent to the LLM. Whistledown
combines pseudonymization and $\epsilon$-local differential privacy ($\epsilon$-LDP) with transformation caching to provide best-effort privacy protection
without sacrificing conversational utility. Whistledown is designed to have low compute and memory overhead, allowing it to be 
deployed directly on a client's device in the case of individual users. For enterprise users, Whistledown is deployed centrally 
within a zero-trust gateway that runs on an enterprise's trusted infrastructure. Whistledown 
requires no changes to the existing APIs of popular LLM providers.
\end{abstract}


\maketitle

\section{Introduction}

Cloud-hosted LLMs from providers like Anthropic~\cite{anthropic2024claude} and OpenAI~\cite{achiam2023gpt} 
have become ubiquitous for general advice, emotional support, and casual conversation. However, 
using cloud-hosted LLMs poses privacy risks when user prompts contain personal or sensitive data 
sent to untrusted servers. This risk is particularly pronounced for what we term \textit{tea-spilling} prompts, 
i.e., conversations where users discuss colleagues, friends, romantic partners, or adversaries by name, 
often seeking advice on interpersonal conflicts or sharing emotionally charged narratives.

The same fundamental challenge is also faced by enterprises and smaller organizations that want to use LLMs for internal communication and decision-making.
For example, an enterprise may want to use LLMs to help employees make decisions about company policies, products, or services. However, 
these prompts may contain sensitive data that the enterprise does not want to share with the LLM. Similarly, an enterprise might want to 
use LLMs to analyze candidate job applications. However, these prompts may contain sensitive data about the applicant that the enterprise 
does not want to share with the LLM. Furthermore, the enterprise may want to minimize demographic bias in the LLM's analysis regarding 
the job applicant.

There are many existing approaches to this problem, a lot of which are not necessarily mutually exclusive. 
The first is to build a user learning program~\cite{merritt2024building}, for example, advising users 
not to share real names. However, user training is error-prone, breaks conversational flow in the LLM context, 
and adds unnecessary cognitive overhead.  A second approach is to only use open-source or in-house LLMs running on a user's device or in an 
enterprise's infrastructure~\cite{deepseek2024}. However, this approach either sacrifices output quality~\cite{nanochat} 
or requires significant compute resources~\cite{deepseek2024}. A third approach to this problem is sanitizing prompts in trusted 
infrastructure before they are sent to an external LLM. This is identical in spirit to the NIST 800-122 recommendation~\cite{mccallister2010guide}
for data sanitization before external release. At its core, Whistledown attempts to solve this problem using the third approach.

\subsection{Deployment Context}

Dorcha provides an AI security platform that exposes a zero-trust gateway located between users and external LLM providers. 
Dorcha's Gateway helps enterprises de-risk the process of integrating their internal infrastructure with LLMs. 
The gateway can be used to manage non-human identities, define secure access policies between internal entities 
and LLM agents, anonymize metadata (e.g., IP addresses, device info), and make all requests between internal entities and agentic services observable and 
auditable. 

Dorcha also provides a lighter-weight tool for everyday users called Saoirse. Saoirse is a client-side AI assistant that sends 
all user requests to a managed instance of Dorcha's Gateway that we run in a trusted execution environment. Saoirse lets everyday users and smaller enterprises benefit 
from common-sense security policies without managing their own infrastructure. Saoirse is essentially an application providing 
a ChatGPT-like conversational interface but with built-in security and privacy functionality.

Recently, users of Saoirse expressed discomfort when including real personal names and other socially relevant sensitive data in prompts. 
This was especially true when users were seeking advice or spilling the tea on people they know in real life. 
Users are concerned that these names could be logged, leaked, or used for training by external models. Beta-users of the Saoirse application
shared that they were manually substituting fake names and other sensitive data to mitigate this risk, which is error-prone and cumbersome.

A similar concern was brought up during discussions with enterprise leaders who wanted to use LLMs for internal communication and decision-making. 
The enterprises work primarily in the humanitarian and non-profit sectors, with employees and users from a diverse range of backgrounds.
The enterprises are interested in making data-driven decisions regarding highly sensitive topics such as human rights violations or 
medical aid distribution. Therefore, integrating LLMs into their decision-making processes, even for casual data gathering or advice, 
comes with significant risks.

To address this, we implemented Whistledown, a best-effort privacy and bias mitigation layer that modifies prompts after a user has submitted them 
but before they are sent to the LLM. Whistledown is designed to have 
very low performance overhead. This allows it to be deployed directly on a client's device in the case of non-enterprise users. When deployed on 
a client's device, Whistledown modifies the prompt before it exits the user's device so only the user sees the unmodified prompt. For enterprise users, 
Whistledown can also be deployed centrally within Dorcha's zero-trust gateway that runs on an enterprise's trusted infrastructure. When deployed within
Dorcha's Gateway, Whistledown modifies the prompt prior to it leaving the enterprise's infrastructure and being sent to a cloud-hosted LLM. This 
allows enterprises to share the prompt unmodified with internal services even with Whistledown enabled and only modifying the prompt at the network 
egress point before it exits the enterprise's internal infrastructure and is sent to a cloud-hosted LLM.

\subsection{Contributions}

We present Whistledown, a client-side privacy system for conversational AI that addresses the unique challenges of protecting personally 
identifiable information (PII) while preserving the utility of multi-turn, emotionally nuanced conversations. Our contributions are:

\noindent\textbf{Conversation Coherence.}
We implement a consistent mapping between plaintext tokens and modified tokens. This mapping helps keep the conversation
coherent and natural from the user's perspective. As an example, 
if a user mentions the name ``Mary'' in a conversation and Whistledown replaces it with ``Lily'' in the first turn of a multi-turn conversation, 
Whistledown will replace all subsequent mentions of ``Mary'' with ``Lily'' in the same conversation. For users that prefer to keep 
aspects of identity in the conversation such as gender, Whistledown's mappings are gender-aware. For instance, ``Mary'' may be mapped to ``Lily''
but not to ``John''. Lastly, Whistledown also ensures that name-like tokens are only replaced when they are actually used in the conversation as names. 
For example, ``So I fell down the hill'' remains unmodified, but ``My friend So fell down the hill'' may be replaced with 
``My friend Lily fell down the hill''. This is because the name ``So'' is not being used as a name in the first sentence, 
but is being used as a name in the second sentence.

\noindent\textbf{Resource-aware Privacy Algorithm.}
There is significant heterogeneity in the hardware resources of individual user devices. Enterprises also have different
SLAs for metrics including Time To First Token (TTFT)~\cite{nvidiabenchmarkingllm} and latency. This makes it difficult to apply a 
one-size-fits-all privacy algorithm for all use cases. We designed a resource-aware privacy algorithm that can fit the needs of both 
user devices and enterprise SLAs. On all devices, Whistledown does light-weight preliminary identification checks for sensitive data composed of Bloom-filter based membership queries and regular expression matching. Whistledown also adds $\epsilon$-local differential privacy to numeric, temporal, 
and contact data fields on all devices. Whistledown additionally requires an on-device BERT~\cite{devlin2018bert} model for context-aware transformations and an on-device Named-Entity-Recognition (NER) model~\cite{zaratiana2024gliner} for context-aware identification. We vary the size of these models based on the resources available on the user's device or an enterprise's TTFT requirements. On the most resource-constrained devices, we disable both models and entirely rely on our preliminary identification checks instead of the NER model and random sampling instead of the BERT model. This provides a better performance-utility trade-off than a one-size-fits-all approach.

\noindent\textbf{Broader Impact.}
This work demonstrates that privacy protection can be deployed on the edge without 
requiring users to trust external infrastructure or sacrifice conversational quality. While we focus on the conversational AI use-case, our techniques generalize to any scenario where users want to share sensitive information with untrusted services while maintaining utility. Examples of broader, non-LLM use-cases include healthcare applications, financial advisors, legal consultation platforms, and more.

\section{Background}\label{sec:background}

\noindent\textbf{LLM Architecture.}
Most modern LLMs are based on the transformer architecture~\cite{vaswani2017attention}. While the original transformer consisted of 
both an encoder and decoder, contemporary LLMs like GPT~\cite{achiam2023gpt} and Claude~\cite{anthropic2024claude} use decoder-only architectures. 
When processing an input prompt, the model first applies a tokenizer to convert text into discrete \textit{tokens}, 
which can be words or subwords. These tokens are then mapped to continuous vector representations called \textit{embeddings} 
through a learned embedding layer. The vector space of these embeddings is called the embedding space. 
The embedding space is a high-dimensional space where semantically similar words tend to be geometrically close to each other, 
as measured by metrics such as cosine similarity~\cite{mikolov2013distributed} or Euclidean distance. The transformer may then processes 
these embeddings through multiple layers of self-attention and feed-forward networks to generate the output response.

\noindent\textbf{Differential Privacy.}
Differential Privacy (DP)~\cite{dwork2006calibrating,dwork2014foundations} provides formal guarantees on the 
amount of information on an individual's data that can be gained from a computation's output. 
A randomized mechanism $M$ satisfies $\epsilon$-DP if for all neighboring datasets $D, D'$ differing in one record, and all outputs $S$:
$$P[M(D) \in S] \leq e^{\epsilon} \cdot P[M(D') \in S]$$
Differential privacy techniques accomplish this guarantee by adding a controlled amount of randomized noise to the data.
This noise is typically drawn from a Laplace distribution, but other distributions can be used as well~\cite{dwork2014foundations}.
The privacy budget $\epsilon$ can be conceptualized as the amount of privacy loss that is acceptable to the user~\cite{snowflakeprivacyloss}. A 
higher $\epsilon$ value means more privacy loss, and a lower $\epsilon$ value means less privacy loss.

The randomized mechanism required for DP can be applied either by a user before sharing their data externally, 
or by a trusted curator that aggregates the data from multiple users before sharing it externally.
These two approaches to differential privacy are called Local Differential Privacy (LDP) and Central Differential Privacy (CDP) respectively~\cite{desfontainesblog20190627}.
CDP is useful in an enterprise setting where it is okay for the data to be accessed within the enterprise's infrastructure, but not externally.
The trusted curator then lies at the network egress point before the data is sent to an external service. 
When Whistledown is deployed within Dorcha's Gateway, it is a CDP mechanism, since the Gateway lies
between the enterprise's internal infrastructure and external LLM services. When Whistledown is deployed directly on a user's 
device running Dorcha's client-facing tool Saoirse, it is an LDP mechanism, since the user's device is the trusted party and 
the data is randomized before it leaves the user's device.

\noindent\textbf{Bloom filters.}
A Bloom filter~\cite{bloom1970space} is a probabilistic data structure that is used to test whether an element is a member of a set. Bloom filters
have one-sided error guarantees, i.e., they have false positives but no false negatives. A false positive, in this case, 
implies that the element is not a member of the set, but the Bloom filter claims it is. A Bloom filter is a bit array of size $m$ 
where each bit is initialized to 0. To add an element $x$ to the encoded set $S$, $x$ is hashed by each of the $k$ hash functions and the 
corresponding bits in the bit array are set to 1. To test whether an element $y$ is a member of the set, $y$ is hashed by each of the $k$ hash functions and the 
bits are checked. If all the bits are 1, then $y$ is a member of the set. If any of the bits are 0, then $y$ is not a member of the set.
The false positive rate of a Bloom filter is approximately $\epsilon = (1 - e^{-kn/m})^k$~\cite{broder2004bloom}. 
The optimal number of hash functions in a Bloom filter is $k = \frac{m}{n} \ln 2$~\cite{broder2004bloom}. By using the optimal value for $k$ 
in the approximation for $\epsilon$, we get $\epsilon = (1 - e^{-\ln 2})^{(\frac{m}{n} \ln 2)} = (\frac{1}{2})^{\frac{m}{n}\ln 2}$. 
Solving for $m$, we get $m = -\frac{n \ln \epsilon}{\ln^2 2}$. For a false positive rate as little as $\epsilon = 0.025$, a 
Bloom filter then requires only about $m/n = -\frac{\ln 0.025}{\ln^2 2} \approx 7.68$ bits per element, or under $1$ Byte per element.

\section{Design}

Whistledown runs as a sidecar service to either our private AI assistant Saoirse or Dorcha's Gateway. 
When it is bundled with Saoirse, it runs directly on a user's device. We will refer to this deployment as \textit{Whistledown-Device}. 
When it is deployed within Dorcha's Gateway, it runs inside the same container as the Gateway on an enterprise's trusted infrastructure. 
We will refer to this deployment as \textit{Whistledown-Gateway}.

\subsection{Whistledown-Device}

\begin{figure*}
    \centering
    \includegraphics[width=0.75\textwidth]{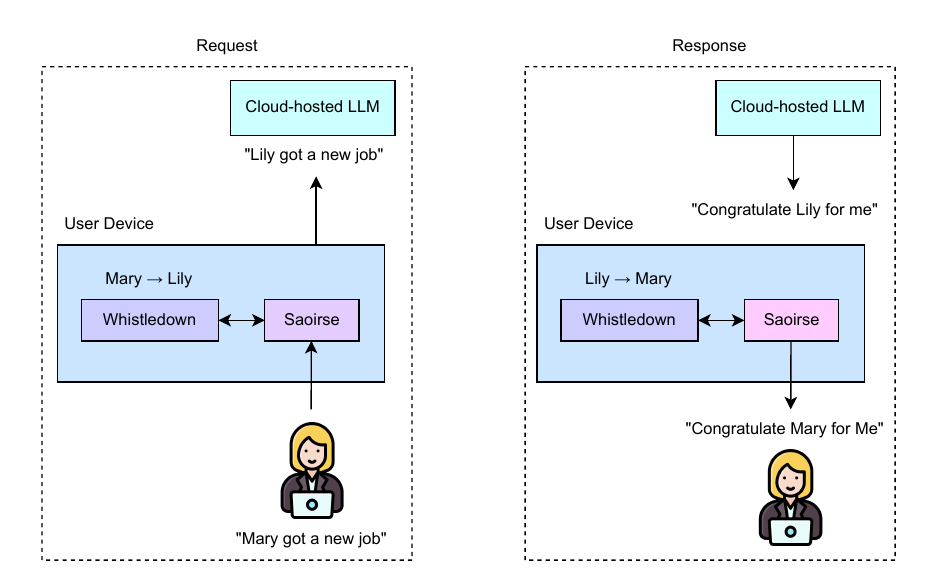}
    \caption{The life of a prompt in a Whistledown-Device deployment. All transformations are performed on a user's device before the prompt is sent to the LLM.
    A consistent mapping between plaintext tokens and modified tokens is maintained to ensure conversational coherence. This ensures that the user's experience
    while conversing with the LLM is not disrupted with Whistledown enabled.}
    \label{fig:whistledown-device}
\end{figure*}

When a user sends a prompt through Saoirse, Whistledown modifies the prompt before it is sent to the LLM. Using Saoirse' UI, the user 
configures the privacy budget ($\epsilon$) and an optional flag for gender debiasing. Saoirse sends these configuration parameters along 
with the user's prompt to Whistledown. Upon receiving the request, Whistledown first identifies sensitive entities in the text. 
For simple fields such as phone numbers, zip-codes, social security numbers, and credit card numbers, Whistledown 
adds $\epsilon$-local Laplace noise to the value based on the user's privacy budget and the global sensitivity of the field. The user
can also choose to simply redact these fields instead of adding noise. The challenge is identifying and transforming more complex fields 
such as names and cities in the prompt, while taking into account the resource constraints of the user's device. For simplicity, 
we will focus on the case of names throughout this section. Other fields are handled similarly.

\noindent\textbf{Context-aware Fields.} We maintain an internal dictionary of names, including multiple spelling variations of the same name, 
that is constantly updated. However this dictionary is already large enough to take up significant main memory on most user devices. 
Loading it from disk noticeably increases the time to first token (TTFT) of the conversation.
Instead, we use a two-stage detection approach. We first encode our internal dictionary of names into a Bloom filter, which is pre-packaged 
with Whistledown. Since we know the size of the dictionary, we can compute the optimal size of the Bloom filter 
at deployment time using the formula derived in Section~\ref{sec:background}. 
We then use the Bloom filter to quickly filter out non-names from the prompt. Since a Bloom filter does not 
have false negatives, as long as our internal dictionary is relatively complete, any tokens that are not in the Bloom filter are guaranteed 
to be non-names.

After this initial filtering, we need to handle two specific types of false positives. 
The first type of false positives are tokens that the Bloom filter identified as names but are not part of our internal dictionary. The second 
type are tokens that are part of our internal dictionary but are not being used as names in the conversation. An example of the second type of 
false positive that we observed is the Korean name ``So'' which is part of our internal dictionary but is not always used as a name in the 
conversation. For this second type of false positive, we need to have a context-aware matching mechanism to distinguish between, for example, 
the sentence ``So I fell down the hill'' and the sentence ``My friend So fell down the hill''. 

To handle these two types of false positives, we pass any tokens that are in the Bloom filter to GLiNER~\cite{zaratiana2024gliner}, 
a light-weight open-source NER model. Since GLiNER is a generalist model, we only need to load 
a single instance of the model in memory to detect all context-aware fields including names, cities, countries, and organizations.
This scales better than loading large datasets for each individual field. More importantly, since GLiNER is only used when our 
initial Bloom filter lets a token through, we can avoid the overhead of running GLiNER on the majority of tokens in the prompt, ensuring 
TTFT remains low.

\noindent\textbf{Transformation.} Once a token is identified as a name, we need to transform it to a pseudonym. 
Whistledown pre-packages a small dictionary of common names. While the dictionary currently stores the most common names by frequency, we are in the process of hand-curating a list of names that is widely believed to be potentially resistant
to name-based discrimination in some societies. For example, the common name ``John''~\cite{brown2023womenceos} will still be included in a future version of our small dictionary, but other common names such as ``Malcolm''~\cite{baldwin1998fire,oyelana2025quantification}, ``Muhammad''~\cite{Said1978,Mamdani2004}, or ``Erin''~\cite{de2004eternal,fried2016no} associated with demographics that have been historically stereotyped might not be.
We replace the original name in the prompt with a random pseudonym from this dictionary. For users that prefer to keep 
aspects of identity in the conversation such as gender, we maintain separate small dictionaries by gender, including a non-binary dictionary.
We use a very light-weight gender classification model trained on our large internal dictionary of names to predict the gender of the original name
in the prompt. We then replace the original name with a random pseudonym from the appropriate gender-specific dictionary. 

For devices with enough resources to run a BERT model, we use a more sophisticated embedding-based LDP mechanism 
to transform the name instead of random replacement from the gender-specific dictionaries. Let $w$ be the original name in the prompt.
The embedding-based LDP mechanism first generates a word embedding $T = \text{BERT}_{\text{TINY}}(w)$ for the name using $\texttt{BERT}_{\text{TINY}}$, 
a lightweight BERT model released by Google~\cite{turc2019well}. $\texttt{BERT}_{\text{TINY}}$ is a 4-layer, 128-dimensional BERT model that is 
trained on the English Wikipedia and BooksCorpus. It is small enough to fit in memory of most user devices and fast enough to not add 
significant latency to the conversation. It is also a good model for this use-case because it is trained on a large corpus of English text, 
including names. This allows $\texttt{BERT}_{\text{TINY}}$ to learn the semantic meaning of names and how they are used in context. 

After generating the word embedding $T$, we add $\epsilon$-local Laplace noise to the embedding to get a noisy embedding $T^{\prime}$.
More precisely, for each element $T_{i}$ of the embedding vector $T$, $T^{\prime}_{i} = T_{i} + Y$, where $Y \getsdollar \text{Lap}(\frac{1}{\epsilon} \cdot 2 \sqrt{d})$. Note that $d$ is the embedding dimension.
We then find the nearest neighbor $T^{\prime\prime}$ of the noisy embedding vector $T^{\prime}$ in the embedding space from 
the small dictionary we pre-packaged with Whistledown for the predicted gender. Finally, we return $w^{\prime} = \text{BERT}_{\text{TINY}}^{-1}(T^{\prime\prime})$.

For users that have enabled gender debiasing, we do not use the embedding-based LDP mechanism, or gender-specific dictionaries. Instead, we 
replace the original name with an identifier of the form ``Person-<uuid>''. This uuid is generated using a cryptographically secure random 
number generator. We also replace the pronouns in the prompt with gender-neutral pronouns. This is done using a simple rule-based approach 
that maintains a static gendered to gender-neutral mapping of all pronouns in the English language. Finally, we replace gendered references 
such as ``husband'' and ``wife'' with gender-neutral references such as ``spouse'', in a best-effort manner.

\noindent\textbf{Caching.} For all transformations, we cache the mapping between the original name and the transformed name in a dictionary. 
This ensures that the same name is always transformed to the same pseudonym within a session, preserving conversational coherence. This cache 
is also used to ensure our transformations are invisible to the user. For example, if a user mentions the name ``John'' in a conversation and Whistledown 
replaces it with ``Matthew'' in the transformed prompt, Whistledown will also replace all mentions of ``Matthew'' in the LLM's response back to ``John''
using the same cache. From the user's perspective, the conversation remains coherent and natural, using the original name throughout.

Figure~\ref{fig:whistledown-device} shows the transformation of a prompt in a Whistledown-Device deployment, both when the user sends it to 
the LLM as well as when the LLM responds back to the user.

\subsection{Whistledown-Gateway}

\begin{figure}
    \centering
    \includegraphics[width=\columnwidth]{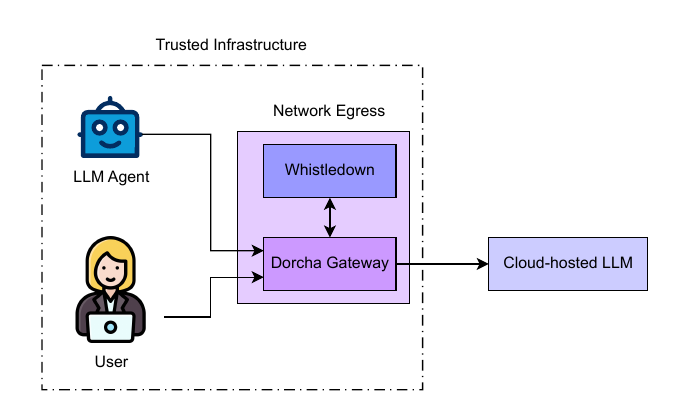}
    \caption{In a Whistledown-Gateway deployment, all transformations are performed on an enterprise's trusted infrastructure before the prompt is sent to the LLM.
    Dorcha Gateway maintains per-user mappings and its users may include non-human identities such as LLM agents and service accounts. The privacy controls are configured centrally by the enterprise.}
    \label{fig:whistledown-gateway}
\end{figure}

Whistledown-Gateway has the same fundamental design as Whistledown-Device, but is deployed within Dorcha's Gateway on an enterprise's trusted 
infrastructure. We restrict our discussion to the parts of the design that are different from Whistledown-Device. Unlike Whistledown-Device, 
Whistledown-Gateway maintains a per user transformation cache and per user privacy budget centrally. This allows enterprises to configure the privacy 
budget for each user and to audit the usage of the privacy budget. Moreover, Whistledown-Gateway allows the enterprise to centrally set a policy 
for many configuration knobs including what fields to transform, what transformation mechanism to use, the privacy budget, and the gender debiasing 
flag. These policies can be set at the granularity of an individual agentic workflow and are updated dynamically without a redeployment of the 
Gateway or any client-side changes.

As an example, an enterprise may want to transform all names in the prompt to pseudonyms and redact all phone numbers and email addresses. 
They may also want to set a gender debiasing policy for workflows that help with employee recruitment and employee performance management 
but not for workflows that help with employee wellness and mental health. Figure~\ref{fig:whistledown-gateway} depicts a Whistledown-Gateway 
deployment.

\section{Examples}

We now demonstrate the use of Whistledown with a few real-world examples. We first use Whistledown without 
embedding-based LDP to transform the following sentence:

\begin{lstlisting}[language=,basicstyle=\ttfamily\small]
I can't believe Ibrahim told the boss that Fatima was slacking off. He's such a snake. Fatima is actually my friend and works really hard.
\end{lstlisting}
The transformed sentence is:

\begin{lstlisting}[language=,basicstyle=\ttfamily\small]
I can't believe Erick told the boss that Licia was slacking off. He's such a snake. Licia is actually my friend and works really hard.
\end{lstlisting}
We now use Whistledown with embedding-based LDP to transform the same sentence, with $\epsilon = 1.0$:

\begin{lstlisting}[language=,basicstyle=\ttfamily\small]
I can't believe Cooper told the boss that Bell was slacking off. He's such a snake. Bell is actually my friend and works really hard.
\end{lstlisting}
However, when we significantly increase the privacy budget to $\epsilon = 400.0$ (i.e. allow for more privacy loss), the transformed sentence is:

\begin{lstlisting}[language=,basicstyle=\ttfamily\small]
I can't believe Yusuf told the boss that Medina was slacking off. He's such a snake. Medina is actually my friend and works really hard.
\end{lstlisting}
Finding an accurate explanation for why, for example, the name ``Yusuf'' is considered closer to ``Ibrahim'' in the embedding space as 
compared to ``Erick'' is beyond the scope of this paper. However, we have noticed multiple times that names that are nearest neighbors in 
the embedding space also share similar demographic attributes in the real world. For example, ``Yusuf'' and ``Ibrahim'' are both Arabic male 
names, and ``Fatima'' and ``Medina'' are both Arabic female names.

We now try a second example, where we transform a sentence that contains a location. We use $\epsilon = 1.0$ and embedding-based LDP for this example:

\begin{lstlisting}[language=,basicstyle=\ttfamily\small]
Sarah and her husband went to Dallas for their anniversary, but she came back alone! Sarah said he had a business emergency in Chicago, but I saw him at a restaurant in Miami with another woman. The whole thing is so suspicious.
\end{lstlisting}
The transformed sentence is:

\begin{lstlisting}[language=,basicstyle=\ttfamily\small]
Meadow and her husband went to Cairo for their anniversary, but she came back alone! Meadow said he had a business emergency in Amman, but I saw him at a restaurant in London with another woman. The whole thing is so suspicious.
\end{lstlisting}
This also illustrates Whistledown's consistent token mapping behavior. For example, ``Sarah'' is mapped to ``Meadow'' in both sentences within the prompt. We now enable gender debiasing for the same sentence, and get the following transformed sentence:

\begin{lstlisting}[language=,basicstyle=\ttfamily\small]
Person-7e6657 and their spouse went to Gujranwala for their anniversary, but they came back alone! Person-7e6657 said they had a business emergency in Shanghai, but I saw them at a restaurant in Ho Chi Minh City with another person. The whole thing is so suspicious.
\end{lstlisting}
Finally, we show an example of a sentence that contains a number of simple fields that are noised using the Laplace mechanism. We use $\epsilon = 1.0$ for this example:

\begin{lstlisting}[language=,basicstyle=\ttfamily\small]
I found out that Robert's credit card 4532-1234-5678-9010 was used to buy expensive gifts for someone named Jessica. When I called him at (555) 987-6543, he denied everything. But then I saw emails between him and jessica.martinez@email.com about meeting up. Something's definitely going on.
\end{lstlisting}
The transformed sentence is:

\begin{lstlisting}[language=,basicstyle=\ttfamily\small]
I found out that Colon's credit card XXXX-XXXX-XXXX-8928 was used to buy expensive gifts for someone named Mia. When I called him at (555) 987-6673, he denied everything. But then I saw emails between him and m***********@email.com about meeting up. Something's definitely going on.
\end{lstlisting}

\section{Analysis}

\subsection{Threat Model}

We consider the external LLM provider as the adversary in our threat model. This model aligns with 
standard assumptions in privacy-preserving systems~\cite{dwork2014foundations} and reflects real-world concerns about 
data logging, model training on user data, and potential data breaches at LLM providers.

\noindent\textbf{Adversary's Goals.} The adversary aims to learn sensitive information from user prompts sent to cloud-hosted LLM services. 
Specifically, the adversary seeks to extract personally identifiable information (PII) such as names, 
locations, phone numbers, and email addresses from user prompts. Beyond direct PII extraction, 
the adversary may attempt to infer demographic attributes including gender, ethnicity, age, and socioeconomic status 
of individuals mentioned in user prompts. The adversary's ultimate goal is to build comprehensive long-term profiles 
of users based on accumulated prompt history, enabling them to track individuals' activities, relationships, locations, 
and interests over time. These learned profiles can then be exploited for unauthorized purposes such as targeted advertising, 
model training on sensitive data, sale to third-party data brokers, or even more malicious activities like identity theft 
or social engineering attacks. The threat is particularly acute given that users often share deeply personal information with LLMs, 
including health concerns, financial situations, workplace conflicts, and family matters, all of which become valuable 
intelligence for adversaries with access to unprotected prompts.

\noindent\textbf{Adversary's Capabilities.} We assume the adversary has complete visibility into all prompts sent to their LLM service, 
including both original prompts from unprotected users and transformed prompts from Whistledown users. The adversary can log and store 
all observed prompts indefinitely, creating a persistent database of user interactions that can be analyzed retrospectively.
We assume the adversary has substantial computational resources and can perform sophisticated statistical analysis on accumulated prompts 
to identify patterns, cluster similar users, and potentially link multiple sessions to the same individual. The adversary may also have access 
to some auxiliary information about users, such as publicly available data from social media, leaked databases, or other external sources. 
However, we assume the adversary does not have comprehensive datasets about all users and cannot directly observe users' true identities 
or non-linguistic metadata beyond what is revealed in the prompts themselves. Note that Dorcha's Gateway strips IP addresses, device fingerprints, 
and other network-level metadata before forwarding prompts to external LLMs.

We explicitly assume the adversary is \textit{honest-but-curious}, i.e., they provide the LLM service correctly and respond to prompts as expected, 
but may analyze and retain all submitted prompts for privacy violations. We assume the adversary does not employ active attacks such as timing analysis, 
traffic fingerprinting, model poisoning to leak information, or adaptive queries designed to reverse-engineer Whistledown's transformation algorithms. 
We also assume the adversary does not attempt to compromise Whistledown's infrastructure or collude with other parties to obtain additional information 
about users.

\noindent\textbf{Trusted Components.} We assume certain components of the system remain trusted and secure throughout operation. 
For Whistledown-Device deployments, we trust the user's personal device where Whistledown runs as a client-side service. 
This trust assumption implies that the device is not compromised by malware, the operating system is secure, and the user maintains 
physical control over the device. For Whistledown-Gateway deployments, we trust the enterprise's internal network infrastructure 
and the specific servers where Dorcha's Gateway is deployed. This includes the assumption that enterprise administrators properly 
configure and maintain security controls, that internal networks are protected by appropriate firewalls and intrusion detection systems, 
and that access to Gateway infrastructure is restricted to authorized personnel.

We also assume benign user intent, i.e., users are using Whistledown to protect their own privacy or the privacy of people 
they mention in prompts, not to circumvent legitimate monitoring, hide malicious activities, or deliberately leak others' sensitive information. 
This trust assumption is critical because users with direct access to Whistledown could potentially attempt to reverse-engineer its 
transformations, intentionally construct adversarial prompts designed to leak information in subtle ways, or share sensitive information 
through covert channels that Whistledown does not analyze. While we acknowledge that some users may act maliciously, 
defending against such insider threats is beyond Whistledown's scope\footnote{However, this is not beyond the scope of Dorcha's Platform 
as a whole.}.

\subsection{Security Goals}

Given the threat model described above, Whistledown aims to provide three primary security guarantees that collectively protect 
user privacy against an honest-but-curious adversary.

\noindent\textbf{Direct PII Protection.} External LLM providers should not be able to observe the plaintext 
values of sensitive fields including names, locations, phone numbers, email addresses, social security numbers, 
credit card numbers, and dates in user prompts. This protection is fundamental because these fields directly identify 
individuals and, if leaked, enable trivial privacy violations such as identity theft, doxxing, or unauthorized contact. 
Whistledown achieves this goal by applying $\epsilon$-local differential privacy transformations and deterministic 
pseudonymization to replace each sensitive field value with a privacy-preserving alternative before prompts leave the trusted domain. 
For complex entities like names and locations, Whistledown employs either dictionary-based pseudonymization by mapping each real name 
to a consistent pseudonym within a session or embedding-based local differential privacy by replacing names with semantically similar 
alternatives drawn from a vocabulary whose selection probability is calibrated by the Laplace mechanism. 
For simple fields like phone numbers and credit card numbers, Whistledown applies format-preserving noise or redaction. 
The key insight is that differential privacy provides a rigorous mathematical guarantee: even with unlimited computational power and 
auxiliary information, the adversary cannot distinguish with high confidence whether a specific individual's information was present 
in the original prompt.

\noindent\textbf{Demographic Privacy.} For users who enable gender debiasing, LLM providers should not be able to infer the 
gender or other demographic attributes of individuals mentioned in prompts from the transformed versions. This goal addresses 
a subtle but important privacy concern. Even if direct identifiers like names are protected, demographic attributes can enable discrimination, 
profiling, or targeted attacks against specific groups. For example, an adversary who learns that a user frequently discusses female colleagues 
in workplace conflict prompts might infer the user's own gender or develop biased models about gender dynamics. 
Whistledown mitigates this threat by replacing gendered names, e.g., ``Sarah'', ``Michael'' with gender-neutral identifiers 
and systematically transforming gendered pronouns, e.g., ``he'', ``she'', to neutral alternatives, e.g., ``they''. 
This transformation preserves the semantic structure and meaning of prompts while removing signals that could reveal protected attributes. 
We note that this feature is optional and user-configurable, as some users may prefer to preserve gender information for tasks where it 
provides important context, such as discussing pregnancy or gender-specific health issues.

\noindent\textbf{Long-term Privacy.} Even if LLM providers log transformed prompts indefinitely and accumulate large datasets 
spanning months or years of user activity, it should be difficult to learn the original sensitive values transformed by Whistledown. This goal partially addresses the risk of long-term data retention and retrospective analysis. Whistledown is helpful here through 
two complementary mechanisms. First, the mathematical guarantees of differential privacy have post-processing immunity. Any additional computation or information combined with DP output cannot worsen the guarantee. Second, Whistledown uses a consistent per-session 
mapping of tokens within a single conversation, ensuring that ``Alice'' always maps to ``Sarah'' throughout a session, 
while randomizing mappings across sessions, so ``Alice'' might map to ``Sarah'' in conversation 1 but ``Jennifer'' in conversation 2.
This helps prevent cross-session linkage so that that even if an adversary observes multiple conversations over time, it is difficult to correlate 
activities or build longitudinal profiles of individuals from the conversations through the transformed fields alone.

\subsection{Open Problems \& Limitations}

While Whistledown attempts to provide strong privacy guarantees within its threat model, it explicitly does not protect against certain categories 
of attacks that fall outside the scope of honest-but-curious adversaries or require fundamentally different system architectures.

\noindent\textbf{Compromised Infrastructure.} If the user's device (for Whistledown-Device deployments) or enterprise infrastructure 
(for Whistledown-Gateway deployments) is compromised by malware, insider attacks, or physical access by adversaries, 
then the adversary gains access to unmodified prompts before transformation occurs. This may be an inherent limitation of any 
client-side or gateway-based privacy system. We cannot protect data that is captured before it enters the trusted computing base. 
Defending against compromised infrastructure requires orthogonal security measures such as endpoint protection, intrusion detection systems, 
secure boot, and trusted execution environments, all of which are beyond Whistledown's scope\footnote{Note that this is a limitation of 
Whistledown specifically, but not of Dorcha's full platform. Dorcha's full platform supports deployment within a trusted-execution 
environment and endpoint protection via client certificates and mTLS, which are beyond the scope of this paper.}.

\noindent\textbf{Malicious Users.} We do not protect against users who deliberately attempt to leak sensitive information about others or 
intentionally circumvent Whistledown's protections. For example, a malicious user could manually transcribe sensitive information 
into prompts in ways that evade detection. Examples include spelling names backwards, using leetspeak, encoding information in semantic patterns, 
or sharing Whistledown's per-session pseudonym mappings with external parties to enable deanonymization. Defending against such adversarial user 
behavior is fundamentally challenging because users have legitimate reasons to access and interact with sensitive information in their prompts. 
Enterprise deployments can partially mitigate this risk through policy enforcement, audit logging, and access controls, 
but it might be very difficult for a technical system to perfectly prevent determined insiders from leaking information they are authorized to see.

\noindent\textbf{LLM-Generated Content.} While Whistledown reverse-transforms names in LLM responses to maintain conversational coherence, 
we do not protect against sensitive information that the LLM generates independently in its responses based on its training data. 
For example, if a user asks ``What are common treatments for diabetes?'', the LLM might respond with general medical information 
that happens to mention real individuals, such as ``Dr. Jane Smith at Johns Hopkins pioneered...''. This leakage originates from 
the LLM itself, not from the user's prompt, and therefore cannot be prevented by Whistledown's input transformations. 
This limitation highlights a broader challenge in LLM privacy. Protecting prompt privacy is necessary but not sufficient for comprehensive privacy, 
as the models themselves may have memorized and can regurgitate sensitive information from their training corpora.

\noindent\textbf{Side-Channel Attacks.} We do not protect against timing attacks, traffic analysis, or other side-channel leakage that 
could be used to fingerprint users, infer information about prompt content, or correlate activities across sessions. 
For example, an adversary who can observe network traffic patterns might infer prompt length, measure Whistledown's processing latency 
to deduce whether embedding-based transformations were applied (which could signal the presence of names or locations), or 
use statistical fingerprinting techniques to link sessions based on typing patterns or conversation structures. 
While Dorcha's Gateway mitigates some of these risks by aggregating traffic from multiple users and stripping IP addresses, 
fully defending against sophisticated side-channel attacks would require constant-time processing, traffic padding, 
and anonymity network integration (e.g., Tor), all of which would impose substantial performance overhead. 

\noindent\textbf{Semantic-based Inference Attacks.} One of the most fundamental limitations of Whistledown is that it does 
not protect against semantic-based inference attacks. This is because Whistledown uses $\epsilon$-local differential privacy 
to protect the direct values of sensitive fields, such as names, phone numbers, dates, etc. This preserves the semantic content 
and context of conversations to maintain utility. However, no amount of noise can prevent an adversary with auxiliary information 
from exploiting this semantic content of the prompt itself to infer the identity of subjects mentioned in transformed prompts. 
For example, consider a user prompt, ``I had a fight with my colleague John about missing the project deadline at Acme Corp.''
Whistledown may transform this to, ``I had a fight with my colleague Michael about missing the project deadline at Beta Inc.'' 
While the names and company are protected via DP-based transformation, an adversary who knows the user works at Acme Corp 
and there was a recent project deadline issue at Acme Corp may infer the user is discussing a specific real colleague, 
despite the name transformation. 

We leave a provable security analysis of Whistledown's threat model and potential ways to mitigate these out-of-scope attacks as open problems for future work.

\section{Implementation \& Evaluation}

\subsection{Implementation}

We have implemented Whistledown as a modular Python service comprising approximately 3.5K lines of code across core transformation modules, 
entity detection components, and API endpoints. The implementation is designed to support both Whistledown-Device and 
Whistledown-Gateway deployment modes through a unified codebase with deployment-specific configuration. We also separately 
wrote a module to integrate Whistledown with Dorcha's Gateway, which is written in Go.

Whistledown exposes a REST API with endpoints for prompt transformation, cache management, and health monitoring. The service is 
containerized using Docker for consistent deployment across edge devices and enterprise gateway infrastructure. For Whistledown-Device deployments, 
we provide a client that communicates with a local service instance. For Whistledown-Gateway deployments, 
the service integrates with Dorcha's Gateway infrastructure and supports policy-based configuration per workflow, 
including $\epsilon$ values, gender debiasing preferences, and resource-aware embedding enablement.

\subsection{Performance Evaluation}

To evaluate Whistledown's runtime overhead, we measure the additional Time To First Token (TTFT) introduced by transformation operations. 
This metric is critical for user experience, as excessive latency would discourage adoption despite privacy benefits. 
We conducted experiments across varying prompt lengths and transformation configurations to characterize Whistledown's performance profile.

\noindent\textbf{Experimental Setup.} We deployed Whistledown on relatively resource-constrained hardware: an M2 Apple MacBook Air with 
8 GB of memory. We tested four sentence configurations: (1) \textit{Simple} sentences containing only base words with no sensitive data, 
(2) \textit{Names} sentences containing person names requiring transformation, 
(3) \textit{Dates} sentences containing date fields to which Whistledown adds Laplace noise, 
and (4) \textit{Names+Dates} sentences containing both entity types. For each configuration, we varied sentence length from 1 to 100 words 
and measured the additional TTFT overhead over 5 iterations, clearing the transformation cache between iterations to ensure cold-start measurements. 
We configured Whistledown with $\epsilon = 1.0$ for all experiments. Critically, since Whistledown parallelizes sentence processing within 
multi-sentence prompts (analyzing each sentence in a separate thread), sentence length rather than total prompt length determines the critical 
path latency for user-perceived TTFT. Our measurements thus capture the per-sentence overhead that directly impacts interactive user experience.
Within a sentence, we interleave a word from each enabled prompt type in a round-robin fashion to create a sentence. 
This means that in a \textit{Simple} sentence, every word is a base word. In a \textit{Names} sentence every other word is a Name. 
In a \textit{Names+Dates} sentence, $\frac{1}{3}$ of the words are Names, $\frac{1}{3}$ of the words are Dates, and the remaining 
$\frac{1}{3}$ of the words are base words.

\begin{figure}[t]
  \centering
  \includegraphics[width=\columnwidth]{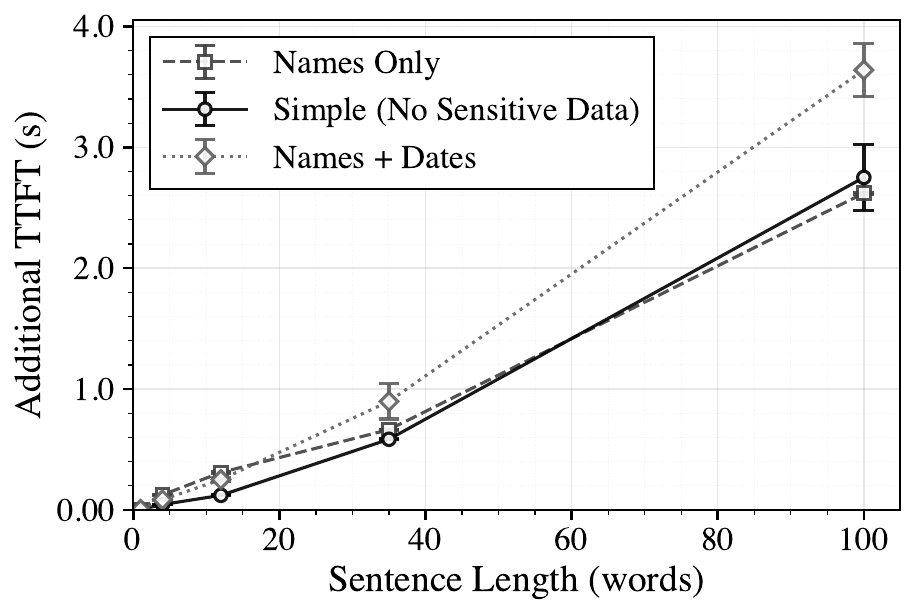}
  \caption{Per-sentence TTFT overhead for Whistledown transformations without embedding-based LDP, across four sentence configurations. Overhead scales approximately linearly with sentence length, ranging from 5-47 ms for 1-2 word sentences to 1.2-1.4 seconds for 59-word sentences. All configurations show similar scaling, indicating entity detection (GLiNER) dominates overhead. Error bars represent standard deviation over 5 iterations.}
  \label{fig:ttft-no-embedding}
\end{figure}

\noindent\textbf{Results.} Figure~\ref{fig:ttft-no-embedding} shows TTFT overhead per sentence for transformations 
without embedding-based LDP (i.e., using dictionary-based pseudonymization for names). For very short sentences (1-2 words), 
all configurations exhibit minimal overhead (5-47 ms), dominated by the fixed cost of invoking the detection pipeline. 
As sentence length increases, overhead scales approximately linearly. Simple prompts reach 121 ms at 12 words, 585 ms at 35 words, 
and 1.2 seconds at 59 words. Prompts containing names show nearly identical scaling, 311 ms at 12 words, 665 ms at 35 words, 
1.3 seconds at 59 words, indicating that dictionary lookup and pseudonym assignment add negligible overhead compared to entity detection. 
Prompts with both names and dates exhibit slightly higher costs due to date transformations requiring Laplace noise generation (252 ms at 12 words, 
898 ms at 35 words, 1.4 seconds at 59 words). The linear scaling reflects GLiNER's sequential token processing. Since typical sentences 
in the English language are under 50 words long~\cite{sigurd2004word}, the overhead remains under 2 seconds per sentence, even when 
$\frac{2}{3}$ of the words require a transformation on the critical path, on relatively resource-constrained hardware. Note that 
this is an evaluation of a preliminary implementation and we will be significantly optimizing the performance of Whistledown in the future 
before rolling it out to production.

\begin{figure}[t]
  \centering
  \includegraphics[width=\columnwidth]{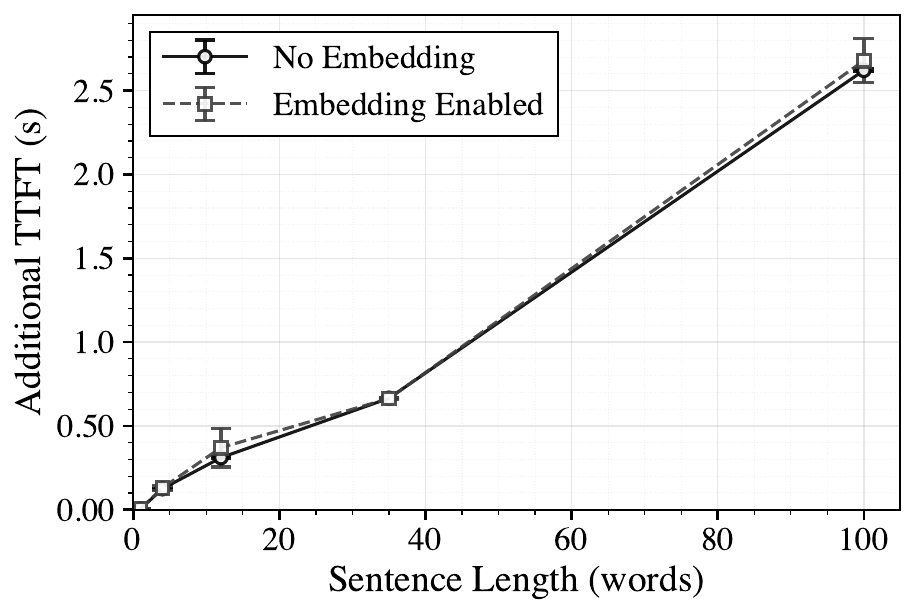}
  \caption{Per-sentence TTFT overhead comparison for name transformations with and without embedding-based LDP. Surprisingly, both approaches show nearly identical overhead across all sentence lengths (within 10\% of each other), indicating that BERT-tiny inference adds negligible cost compared to GLiNER-based entity detection. This suggests embedding-based LDP is practical even for latency-sensitive applications. Error bars represent standard deviation over 5 iterations.}
  \label{fig:ttft-embedding}
\end{figure}

Figure~\ref{fig:ttft-embedding} compares TTFT overhead per sentence for name transformations with and without embedding-based LDP. Surprisingly,
embedding-based transformations show nearly identical overhead to dictionary-based approaches across all sentence lengths. This result 
indicates that $\text{BERT}_{\text{TINY}}$ inference for computing word embeddings and applying the exponential mechanism adds negligible overhead compared 
to the dominant cost of GLiNER-based entity detection. We attribute this to two factors: (1) $\text{BERT}_{\text{TINY}}$ is highly optimized and processes
embeddings only for detected names rather than all tokens, and (2) GLiNER's context-aware NER over the full sentence dominates the computational 
budget. Future optimizations focus on accelerating GLiNER inference.

\section{Conclusion}

We have presented Whistledown, a privacy-preserving transformation layer for cloud-hosted LLM interactions that addresses the growing tension between conversational utility and data privacy. Our work demonstrates that rigorous privacy protections can be deployed transparently without requiring users to modify their behavior, trust external infrastructure, or sacrifice conversational coherence. Whistledown combines $\epsilon$-local differential privacy with 
consistent transformation mapping and resource-aware transformation strategies to provide formal privacy guarantees while maintaining the natural flow of multi-turn conversations. Beyond conversational AI, Whistledown's approach generalizes to any scenario where users want to share sensitive information with untrusted services while maintaining utility. Healthcare applications, financial advisors, legal consultation platforms, and social services all face similar privacy challenges 
that could benefit from transparent, best-effort privacy transformations. As LLMs become increasingly integrated into everyday life and critical infrastructure, systems like Whistledown provide a practical path toward privacy-preserving AI that respects user autonomy 
while enabling the benefits of cloud-scale intelligence.

\bibliographystyle{ACM-Reference-Format}
\bibliography{references}

\subsection*{Generative AI Disclosure}

The paper was written directly by the authors. The authors used ChatGPT by OpenAI and Claude by Anthropic for feedback on technical writing and to edit and debug the matplotlib figure code in the \LaTeX{} manuscript. The artifacts of our work were primarily written and verified by the authors. The second author, HT, used Cursor IDE, which includes a generative AI code assistant, for the following tertiary coding tasks: unit-test generation, code refactoring, feedback on bug-fixing, and code reviews.

\end{document}